\documentclass[pss,fleqn]{w-art}
\usepackage{times}
\usepackage{w-thm}
\usepackage{cite}
\usepackage[]{graphicx}
\newcommand{\figwidth}{\textwidth}
\newcommand{\figwidthhalf}{0.45\textwidth}
\newcommand{\figwidththreed}{0.5\textwidth}

\renewcommand{\vec}[1]{\ensuremath{{\mathbf{#1}}}}
\newcommand{\muB}{\mu_{\rm B}}
\newcommand{\VD}{V_{\rm I}}
\newcommand{\VI}{V_{\rm C}}

\renewcommand{\i}{\ensuremath{{i}}}

\renewcommand{\r}{\ensuremath{\vec{r}}}

\newcommand{\q}{\ensuremath{\vec{q}}}

\newcommand{\dr}{\ensuremath{d^2\vec{r}}}
\newcommand{\Tr}{\ensuremath{\mbox{Tr}}}

\newcommand{\LDOS}{\mbox{LDOS}}

\begin{document}
\DOIsuffix{theDOIsuffix}
\Volume{XX}
\Issue{1}
\Month{01}
\Year{2003}
\pagespan{3}{}
\Receiveddate{\today, compiled $Revision: 1.20 $}
\Reviseddate{30 November 2003}
\Accepteddate{2 December 2003}
\Dateposted{3 December 2003}



\title[Hartree-Fock Interactions in the IQHE]{Hartree-Fock Interactions in the Integer Quantum Hall Effect}


\author[RA R\"{o}mer]{Rudolf A. R\"{o}mer\footnote{Corresponding
     author: e-mail: {\sf r.roemer@warwick.ac.uk}, Phone: +44\,2476\,574\,328,
     Fax: +44\,7876\,858\,246}} \address[]{Department of Physics and Centre for Scientific Computing, University of Warwick, Coventry, CV4 7AL, United Kingdom}
\author[C Sohrmann]{Christoph Sohrmann\footnote{E-Mail: {\sf c.sohrmann@warwick.ac.uk}}}
\begin{abstract}
  We report on numerical studies into the
  interplay of disorder and electron-electron interactions
  within the integer quantum Hall regime, where
  the presence of a strong magnetic field
  and two-dimensional confinement of the
  electronic system profoundly affects thermodynamic and
  transport properties.
  We emphasise the behaviour of the electronic
  compressibility, the local density of states, and
  the Kubo conductivity.
  Our treatment of the electron-electron interactions relies
  on the Hartree-Fock approximation so as to achieve
  system sizes comparable to experimental situations.
  Our results clearly exhibit manifestations of various interaction-mediated
  features, such as non-linear screening,
  local charging, and $g$-factor enhancement, implying
  the inadequacy of independent-particle models for comparison
  with experimental results.
\end{abstract}
\maketitle                   






\section{Introduction}
\label{sec-introduction}

The {\em integer}\ quantum Hall effect (IQHE) --- observed in
two-dimensional electron systems (2DES) subject to a strong
perpendicular magnetic field \cite{KliDP80} --- has been well studied
using single-particle arguments
\cite{Pra81,ChaP95,JanVFH94,Lau81,Pru84,ThoKNN82,Pru87,ChaC88,CaiR04,
KraOK05}.
However, experiments on mesoscopic MOSFET devices have questioned
the validity of such a simple single-particle picture. Measurements of
the Hall conductance as a function of magnetic field $B$ and gate
voltage \cite{CobBF99} exhibited regular patterns along integer filling
factors. It was argued that these patterns should be attributed to
Coulomb blockade effects.
Similar patterns have been found recently also in measurements of the
electronic compressibility $\kappa$ as a function of $B$ and electron
density $n_{\rm e}$ in the IQHE \cite{IlaMTS04} as well as the
fractional quantum Hall effect (FQHE) \cite{MarIVS04}. From these
measurements it turns out that deep in the localised regime between two
Landau levels, stripes of constant width with particularly small
$\kappa$ can be identified. These stripes consist of a collection of
small-$\kappa$ lines, identifiable with localised states, and their
number is independent of $B$. This is inconsistent with a
single-particle picture where one expects a fan-diagram of lines
emanating from $(0,0)$ in the $(B,n_{\rm e})$-plane. The authors of \cite{IlaMTS04,MarIVS04} argue
that their results may be explained qualitatively by non-linear
screening of the impurity charge density at the Landau level band
edges. Clearly, such screening effects --- described within a
Thomas-Fermi approach \cite{IlaMTS04} --- are beyond the scope of a
non-interacting theory. This immediately raises a question on the
status of the aforementioned universality of the QH transition
\cite{CobBF99,ShaHLT98,BalMB98}, which was obtained largely within a
single-particle approach
\cite{AokA85,ChaC88,ChaD88,HucK90,LiuS94,Huc95,SinMG00,CaiR05}. Indeed
a recently developed theory of the IQHE for short-ranged disorder and
Coulomb interactions shows that universality is retained due to
$\mathcal{F}$ invariance \cite{PruB05a,PruB05b,PruB05c}.

In the present paper, we quantitatively investigate the effects of
Coulomb interactions on the compressibility in the $(B,n_{\rm
e})$-plane within a Hartree-Fock (HF) approach. HF accounts for
Thomas-Fermi screening effects while at the same time leading to a
critical exponent $\nu$ whose value is consistent with the results
of the non-interacting approaches \cite{LeeW96,YanMH95,SohR07}. We
find that the observed charging lines in the compressibility can be
well reproduced and that the width of each group of lines is well
estimated by a force balance argument. Thus we can quantitatively
explain the lines and stripes in the compressibility as a function
of $(B,n_{\rm e})$, fully in support of the qualitative picture proposed
in \cite{IlaMTS04,PerC05,SohR05,StrK06}. However, our results for
the conductivity patterns of Ref.\ \cite{CobBF99} do not seem to
reproduce the observed stripes. This might be due to our still too
small system sizes \cite{DiaAPW07} or many-particle physics beyond
the HF approximation \cite{AndFS82}.

\section{The Hartree-Fock approach in Landau basis}
\label{sec-hf}

We consider a 2DES in the $(x,y)$-plane subject to a perpendicular
magnetic
field $\vec{B} = B\vec{e}_z$. The system can be described by a
Hamiltonian of the form
\begin{equation}
H^\sigma_{\rm 2DES} =
           h^\sigma + \VI =
           \frac{(\vec{p}-e\vec{A})^2}{2m^*} +
           \frac{\sigma g^* \muB B}{2} +
           \VD(\vec{r}) +
           \VI(\vec{r},\vec{r}'),
\label{eq-hamiltonian}
\end{equation}
where $\sigma = \pm 1$ is a spin degree of freedom, $\VD$ is a smooth
random potential modelling the effect of the electron-impurity
interaction, $\VI$ represents the electron-electron interaction term and
$m^*$, $g^*$, and $\mu_{\rm B}$ are the effective electron mass,
$g$-factor, and Bohr magneton, respectively \cite{Soh07}.

\begin{figure}[t]
\includegraphics[width=0.85\figwidth]{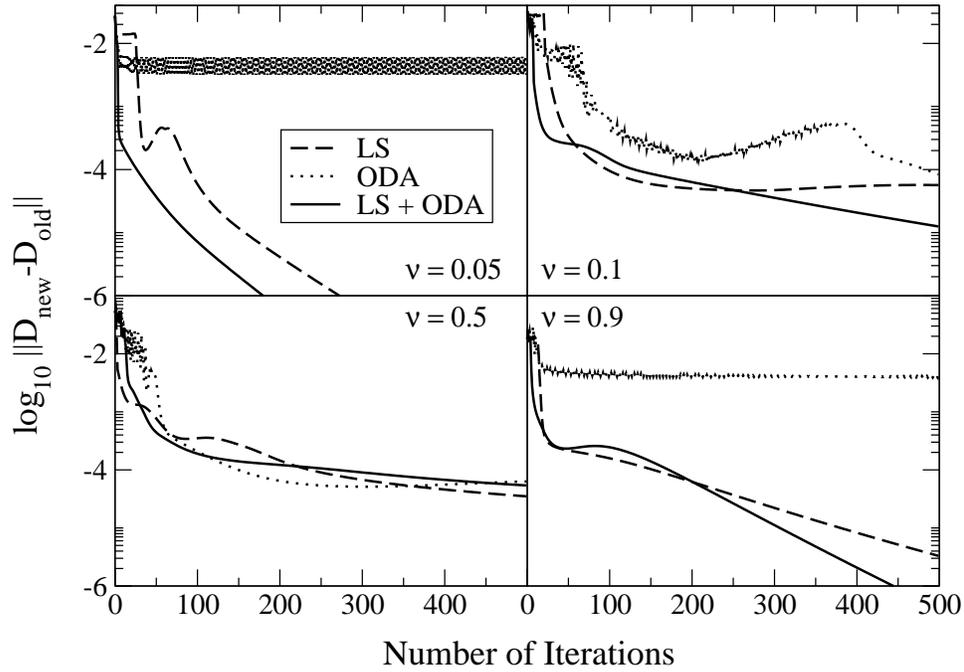}
\caption{Dependence of the convergence on
the number of HF iterations for the three HF algorithms. The four
panels correspond to $4$ different filling factors.}
\label{fig-convergence}
\end{figure}

The electron-electron interaction potential has the form
\begin{equation}
\VI(\vec{r},\vec{r}') =
  \frac{e^2}{4\pi\epsilon\epsilon_0}
    \frac{\gamma}{|\vec{r}-\vec{r}'|}=
  \sum_{\vec{q}} \VI(\vec{q})
  \exp\left[\i\vec{q}\cdot(\vec{r}-\vec{r}')\right],
\label{eq-coulomb-interaction-term}
\end{equation}
with
\begin{equation}
\VI(\vec{q}) = \frac{e^2}{4\pi\epsilon\epsilon_0 l_{\rm
c}}\frac{\gamma}{N_\phi |\vec{q}|l_{\rm c}}.
\end{equation}
The parameter $\gamma$ will allow us to continually adjust the
interaction strength; $\gamma=1$ corresponds to the bare Coulomb
interaction. Using $\gamma<1$ merely serves as a numerical handle
to the interaction strength and might be easily eliminated by
rescaling of the disorder strength. $N_{\phi}=L^2/2\pi l_{c}^{2}$ is the number of states
per Landau level, also referred to as the number of magnetic flux
quanta. Choosing the vector potential in Landau gauge, $\vec{A} =
Bx\vec{e}_y$, the kinetic part of the Hamiltonian is diagonal in the
Landau functions \cite{LanL81}
\begin{equation}
\varphi_{n,k}(\vec{r}) = \langle\vec{r}|nk\rangle =
\frac{1}{\sqrt{2^n n! \sqrt{\pi} l_{\rm c} L}}
\exp \left[\i k y-\frac{(x-kl_{\rm c}^2)^2}{2l_{\rm c}^2}\right] H_n\left(\frac{x-kl_{\rm c}^2}{l_{\rm c}}\right),
\end{equation}
where $n$ labels the Landau level index, $k={2\pi j}/{L}$ with $j= 0,
\ldots, N_{\phi}-1$ labelling the momentum, $H_{n}(x)$ is the $n$th
Hermite polynomial, and $l_{\rm c} = \sqrt{\hbar/eB}$ the magnetic
length. We use $N_{\rm I}$ Gaussian-type "impurities", randomly
distributed at $\vec{r}_s$, with random strengths $w_s \in [-W,W]$,
and a fixed width $d$, such that $ \VD(\vec{r}) = \sum_{s=1}^{N_{\rm
I}} \left(w_s/\pi
  d^2\right) \exp[-(\vec{r}-\vec{r}_s)^2/d^2] = \sum_{\vec{q}}
\VD(\vec{q}) \exp(\i\vec{q}\cdot\vec{r}) $ with
\begin{equation}
\VD(\vec{q}) = \sum_{s=1}^{N_{\rm I}} \frac{w_s}{L^2} \exp
\left(-\frac{d^2|\vec{q}|^2}{4}-\i\vec{q}\cdot\vec{r}_s\right),
\end{equation}
where $q_{x,y} = {2\pi j}/{L}$ and $j= -N_\phi, -N_{\phi}-1, \ldots,
N_\phi$. The areal density of impurities therefore is given by $n_{\rm
  I} = N_{\rm I}/L^2$.
The
Hamiltonian is represented in matrix form using the periodic Landau states
$|nk\rangle$ and we have
\begin{eqnarray}
\mathbf{H}^\sigma_{n,k;n',k'}
&= &\langle nk |H^\sigma_{\rm 2DES}|n'k'\rangle \nonumber \\
&= &\left(n+\frac{1}{2}+\frac{\sigma g^*}{4}\frac{m^*}{m_{\rm e}}\right)
     \hbar\omega_{\rm c}\delta_{n,n'}\delta_{k,k'} +
 \mathbf{V}_{n,k;n',k'} + \mathbf{F}^\sigma_{n,k;n',k'}\quad,
\end{eqnarray}
with the cyclotron energy $\hbar\omega_{\rm c} = \hbar eB/m^*$. The
disorder matrix elements are given by $\mathbf{V}_{n,k;n',k'} =
\sum_{\vec{q}} \VD(\vec{q}) S_{n,k;n',k'}(\vec{q})$, where mixing of
Landau levels is included. The explicit form of the plane wave
matrix elements $S_{n,k;n',k'}(\vec{q}) = \langle nk
|\exp(\i\vec{q}\cdot{\vec{r}})|n'k'\rangle$ and the Fock matrix $\mathbf{F}^\sigma_{n,k;n',k'}$ can be
found in Ref.\ \cite{SohR07}. The total energy $E_{\rm
  tot}$ in terms of the above matrices is given as
\begin{equation}
E_{\rm tot}
             =\Tr(\mathbf{h}\mathbf{D}+\frac{1}{2}\mathbf{F}\mathbf{D})
             = \frac{1}{2}\sum_\sigma \sum_{n,k;n',k'}
             \left(2
\mathbf{h}^\sigma_{n,k;n',k'}+\mathbf{F}^\sigma_{n,k;n',k'}\right)
             \mathbf{D}^\sigma_{n',k';n,k}
\label{eqn-diffEtot}
\end{equation}
where the density matrix is $ \mathbf{D}_{n,a;m,b} = \sum_{\alpha}
f_\alpha C^\alpha_{n,a} C^{\alpha,*}_{m,b} $ with $f_{\alpha}$ the
Fermi function and $C_{n,a}^\alpha$ the expansion coefficients for
the Landau basis \cite{SohR07}. The sums $n$ and $n'$ run over the
number of Landau levels, $N_{\rm LL}$, taken into account. A
variational minimisation of $\langle\Psi|H_{\rm 2DES}|\Psi\rangle$
with respect to the coefficients yields the Hartree-Fock-Roothaan
equation \cite{Roo51}. The simplest scheme for solving this
self-consistent eigenvalue problem is the Roothaan algorithm
\cite{Roo51,CanB00}. However, convergence of the Roothaan algorithm
is rather poor. In most cases it runs into an oscillating limit
cycle \cite{Soh07}. This limit cycle can be avoided be minimising a
penalised energy functional
\begin{align}
E'(\mathbf{D}_{\rm old},\mathbf{D}_{\rm new}) = E(\mathbf{D}_{\rm old},
\mathbf{D}_{\rm new}) + b||\mathbf{D}_{\rm old}-\mathbf{D}_{\rm
new}||^2~,
\label{eq-convergence}
\end{align}
instead of the actual HF energy functional where ``old'' and ``new``
denote two consecutive states in the HF-selfconsistency cycle. This
then leads to the Level-Shifting algorithm (LSA) \cite{SauH73}.
However, while LSA avoids limit cycles, it can and does lead to
unphysical solutions. A further improvement is based on the
Optimal-Damping algorithm (ODA) \cite{CanB00}. The iteration is
carried out just as in the Roothaan algorithm, only that the new
density matrix is a mixture of the old and the new density matrix,
i.e.
\begin{align}
\mathbf{D} = (1-\lambda) \mathbf{D}_{\rm old} + \lambda \mathbf{D}_{\rm
new}~,
\end{align}
with a damping parameter $\lambda$ which is chosen optimally
according to the direction of steepest descent in the total HF energy
\cite{Soh07}.
As it turns out the performance also depends strongly on the
interaction strength and the position of the Fermi level. For some
filling factors and choices of parameters, we might find fast
convergence with one of the algorithms described above. However,
over the whole range of filling factors only a combination of ODA
and LSA can guarantee convergence in any case. In Fig.\
\ref{fig-convergence} we have depicted the convergence behaviour for
a typical HF run at four different filling factors. Note that we
have plotted the convergence  as defined by Eq.\
(\ref{eq-convergence}), rather than the total HF energy, which is
also sometimes used as a measure. In comparison to the energy which
acquires a finite value at convergence, the error estimate
$||\mathbf{D}_{\rm new} - \mathbf{D}_{\rm old}||$ goes to zero and
its convergence is therefore easier to interpret.

\section{Wave function and charge distributions}
\label{sec-distributions}

\subsection{The local density of states}
\label{sec-ldos}

The local density of states (LDOS) is a very interesting quantity
which is experimentally accessible via spatial scanning tunneling
spectroscopy (STS) \cite{DomSWM99,MorKMG02,HasSMI07}. The
differential tunneling current between the STS tip and the sample is
thereby proportional to the density of existing states at a certain
energy. By definition the LDOS is the tunneling density of states (TDOS)
weighted with the charge
density at a position $\r$, defined by
\begin{equation}
\LDOS(E,\r) = \sum_\alpha |\psi_\alpha(\r)|^2 \delta(E-E_\alpha)~,
\end{equation}
such that $\int \dr \LDOS(E,\r) = \sum_\alpha \delta(E-E_\alpha) =
\rho(E)$, where $\rho(E)$ is the TDOS.  In practice, the
$\delta$-function is broadened by temperature as well as an AC
modulation voltage applied when using a lock-in technique for noise
reduction. The energy window can then be described by a semi-ellipse
around $E$ with the broadening $\Delta E$, which is usually of the
order of $1$meV \cite{MorKMG02}. Thus, the LDOS can be regarded as
the charge density in a small energy interval and allows to study
the charge distribution in the 3 dimensions (2 spatial and energy).
\begin{figure}[tb]
  \center
  \includegraphics[width=\figwidththreed]{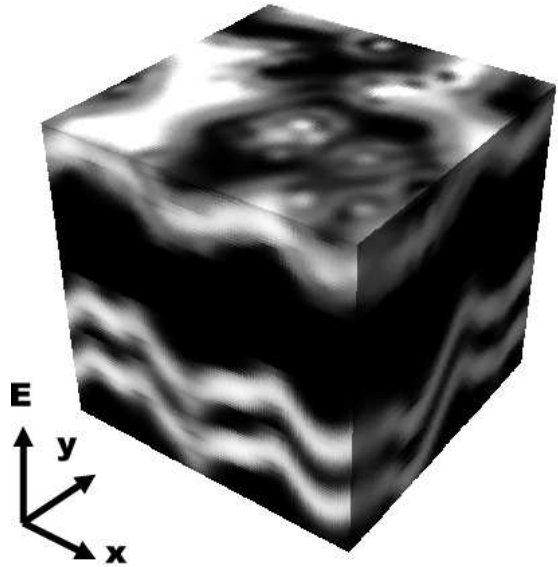}
  \caption{Local density of states for a non-interacting Hall system as a function of $(x,y)$ position as well as energy $E$ for a system of size $L=300nm$. The filling factor $\nu \approx 3$ and $B=5$T. Large LDOS values are indicated by white.}
  \label{fig-ldos}
\end{figure}

\begin{figure}[tb]
  \center
  \includegraphics[width=\figwidthhalf]{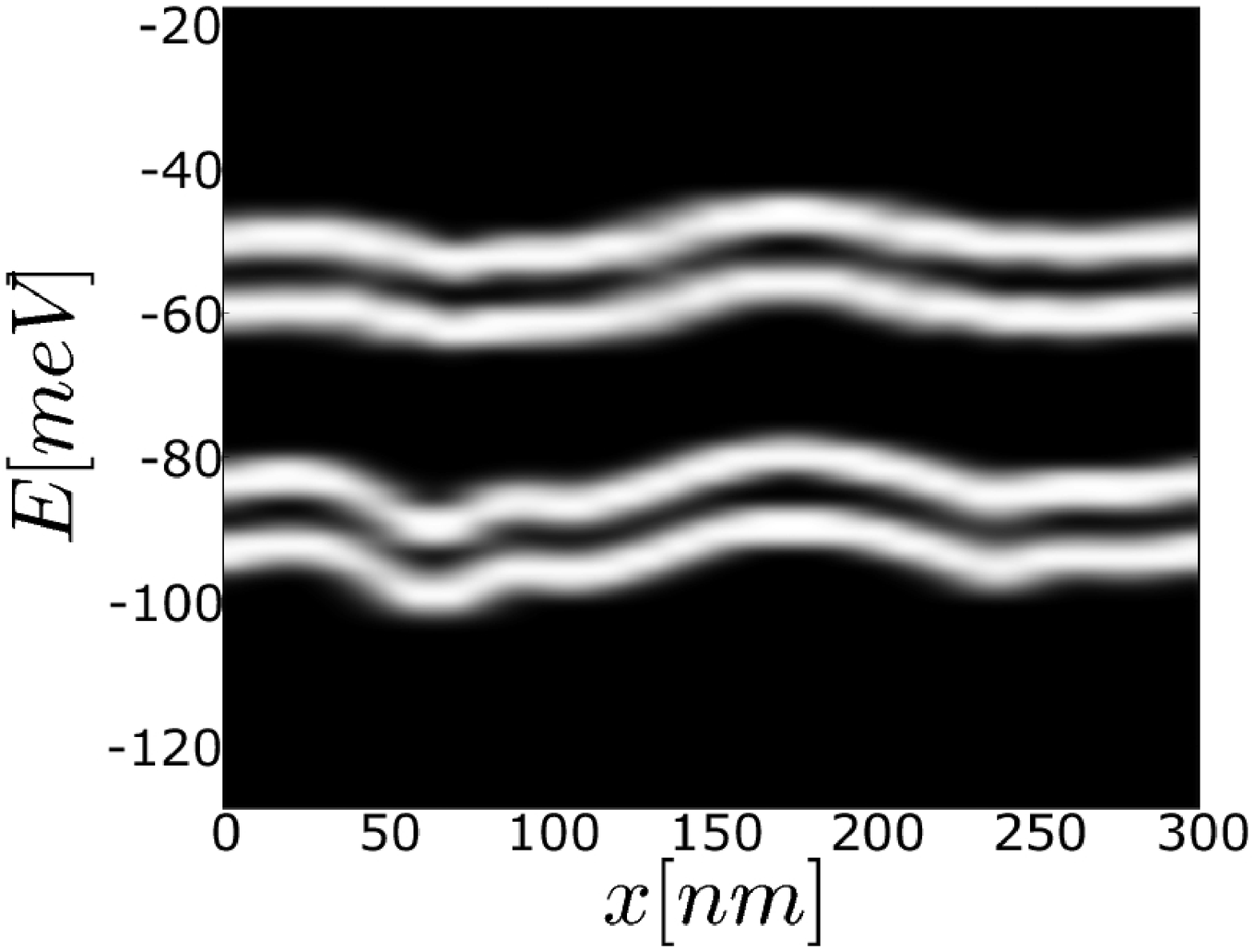}
    \includegraphics[width=\figwidthhalf]{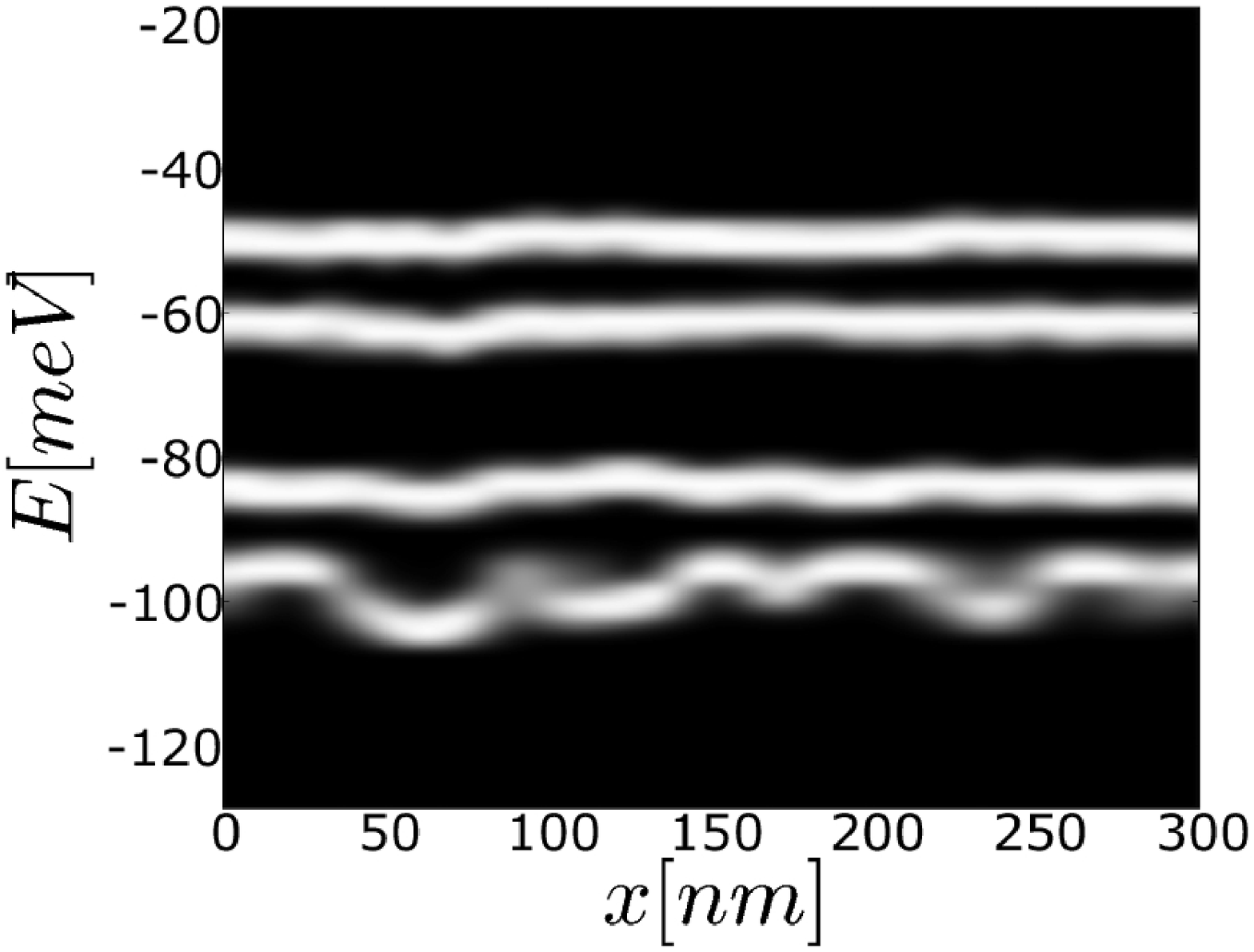}
  \caption{Local density of states as a function of $x$ position as
well as energy $E$ for a system of size $L=300nm$ at constant $y$ and $B=5$T. The filling factor $\nu = 0$ (left) and $\nu =0.5$ (right). Large LDOS values are indicated by white.} \label{fig-ldos-int}
\end{figure}

In Figs.\ \ref{fig-ldos} and \ref{fig-ldos-int} we show results for
a model of an n-type InSb structure of three-dimensionally
distributed donor atoms with a 2DES sitting on top of the cleaved
surface \cite{HasSMI07}. The first $2$ spin-split Landau levels
are shown for Fig.\ \ref{fig-ldos-int}. At filling factor $\nu=0$,
electrons are tunneling from the tip into unoccupied states of the
sample and one thus probes non-interacting physics. A spin-splitting
equal to the Zeeman energy can be seen which is of equal magnitude
in each level.
In the right panel of Fig.\ \ref{fig-ldos-int}, we show the situation at
$\nu =0.5$, i.e. when the Fermi energy lies right in the centre of the lowest Landau level.
We observe that the high-LDOS features of this level appear to be
pushed away from the
Fermi energy, resulting in the onset of the famous Coulomb gap
("Efros-Shklovskii gap", see \cite{EfrS75}) in the TDOS. We would like to note
that the gap in the TDOS shall not be confused with the gaps appearing
in the quasi-particle spectrum. In fact, it has been shown that
despite the linear Coulomb gap in the TDOS, the thermodynamic density of states
remains finite at the Fermi level \cite{YanWM01,SohR07}.
Furthermore, the fluctuations in this level are much stronger than
in the other three levels. We interpret this as a signature of the
screening mechanism, i.e. the electrons near the Fermi level screen
the disorder such that the states in the higher levels only see a
weakened disorder potential. We also see that the spin-splitting in
the first level is enhanced as compared to the non-interacting case.
Similarly, the spin-splitting between levels $3$ and $4$ is also
larger. These effects seems to be related to the direct and indirect
spin splitting enhancements for instance recently reported in
Ref.~\cite{DiaAPW07}.

\subsection{The charge distribution}
\label{sec-chargedistribution}

\begin{figure}[tbhp]
  \center
\includegraphics[width=\figwidth]{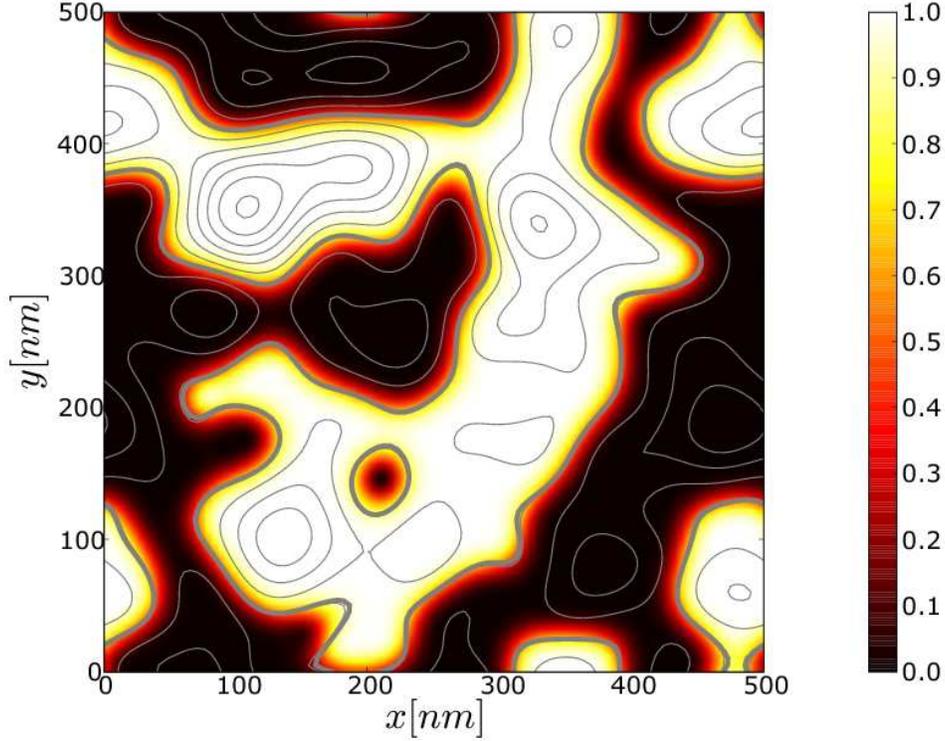}
  \caption
   [Spatial distribution of non-interacting electron density at
$\nu=1/2$]
   {Spatial distribution of non-interacting electron density
    $n(\vec{r})/n_0$ at $B=4$T, $\gamma=0$ and $\nu=1/2$ as indicated by
the
    colour scale. Solid contour lines show the equipotential lines of the
    $\VD(\vec{r})$. The thick solid lines corresponds to $\epsilon_{\rm
F}$. }
  \label{fig-density-non}
\end{figure}

\begin{figure}[tbhp]
  \center

\includegraphics[width=\figwidth]{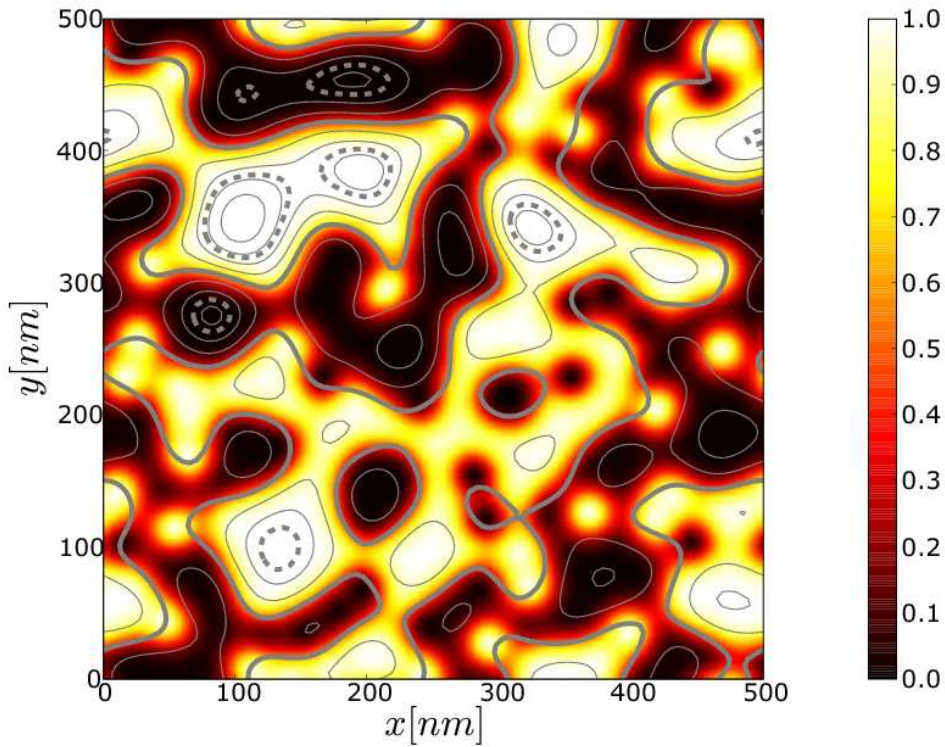}
  \caption
   [Spatial distribution of HF-interacting electron density at
   $\nu=1/2$]
   {Spatial distribution of HF-interacting electron density
    $n(\vec{r})/n_0$ at $B=4$T, $\gamma=0.3$ and $\nu=1/2$ as indicated
by the
    colour scale. Contour lines show equidensity lines of (\ref{eq-quasi-laplace}). The broken lines
    indicate unscreenable (insulating) regions. The thick solid line
shows
    $n_{\rm scr}(\vec{r}) = \bar{n}_{\rm scr} = 0$.}
  \label{fig-density-int-linear}
\end{figure}

The spatial distribution of the total electronic density
\begin{equation}
n(\vec{r})
=\sum_{\sigma}\sum_{\alpha=1}^{M}
\left|\psi_{\alpha}^{\sigma}(\vec{r})\right|^2
= L^{-2}\sum_\sigma \sum_{n,k,n',k'}\sum_{\vec{q}}
\mathbf{D}^{\sigma}_{n,k;n',k'} S_{n,k;n',k'}(\vec{q})
   \exp(-\i\q\r)
\end{equation} 
is also readily calculated in our approach. It details the screening
mechanism by providing direct insight into the interplay of disorder
and interaction. Let us start at the QH transition.
Fig.\  \ref{fig-density-non} depicts the critical charge density at $\nu
= 1/2$ for
a non-interacting system in units of $n_0$.
The contour lines show the impurity potential $V_{\rm I}(\vec{r})$ where
the critical energy $V_{\rm I}(\vec{r}) = \epsilon_{\rm F}$ is
highlighted by a thick line.  The charge density evidently behaves
according to the semiclassical approximation \cite{Huc95} and follows
the equipotential lines of $V_{\rm I}(\vec{r})$.
For the interacting case, however, we expect Thomas-Fermi screening
theory to apply \cite{Efr88a,Efr88b,Efr89a,Efr92}. This
approximation is appropriate for an impurity potential smooth on the
scale of the magnetic length as well as a sufficient separation of
the Landau bands, characterised by the condition $\hbar\omega_{\rm
c}/l_{\rm c} > \sqrt{\langle|\nabla V_{\rm I}(\vec{r})|^2\rangle}$.
The induced electrostatic potential of the charge density
\begin{equation}
\phi(\vec{r}) = \frac{e}{4\pi\epsilon\epsilon_0}\int d^2\vec{r}'
\frac{n(\vec{r}') - \bar{n}}{|\vec{r}'-\vec{r}|}
\end{equation}
and the impurity potential $V_{\rm I}(\vec{r})$ form a screened
potential $ V_{\rm scr}(\vec{r}) = V_{\rm I}(\vec{r}) + e
\phi(\vec{r})$.  Here, $\bar{n}$ accounts for the positive
background. Since a flat screened potential is energetically most
favourable, one expects to find $V_{\rm scr}(\vec{r}) =
\epsilon_{\rm F}$ for the case of perfect screening. However, since
fluctuations of the charge density, $n(\r)$,
are restricted between an empty and a full Landau level,
i.e.~$(N_{\rm LL}-1) n_0 < n(\vec{r}) < N_{\rm LL} n_0$, where
$n_0=(2\pi l_c^2)^{-1}$ is the density of a full Landau level, the
screening is not always perfect but depends on the fluctuations in
the impurity potential as well as on the filling factor
\cite{Efr88a,Efr88b,Efr89a}. With the condition that $\hbar\omega_c$
is large compared to the potential energy scale, we can restrict the
discussion to a single Landau level, i.e.~$N_{\rm LL} = 1$. The
plane can be divided into fully electron or hole depleted,
insulating regions
--- where $n(\vec{r}) = 0$ or $n(\vec{r}) = n_0$, respectively ---
and metallic regions
--- where $n(\vec{r})$ lies in between. Depending on the filling
factor, the extent of those regions varies. Close to the band edge,
insulating regions dominate. Screening is highly non-linear and
transport virtually impossible. On the other hand, if disorder is
weak enough, there exists a finite range of filling factors in the
centre of each band where metallic regions cover most of the sample,
percolate and render the whole system metallic. The disorder is
effectively screened and transport greatly enhanced. In that case,
the charge density $n_{\rm
  scr}(\vec{r})$ can be obtained by Fourier transforming the screened
potential. In 3D, this simply leads to the Laplace equation. For 2D \cite{WulGG88},
however, one obtains
$
n_{\rm scr}(\vec{q}) = -(2\epsilon\epsilon_0/e^2)|\vec{q}|V_{\rm
I}(\vec{q}) + \nu n_0 \delta_{q,0}, 
$
where the $|\vec{q}| = 0$ only couples to the homogeneous, positive
background and thus does not contribute to screening of the impurity
potential. In other words, in our model only the fluctuations
$\delta n(\vec{\r})$ are essential for screening.
Hence, in 2D, a perfectly screening charge density would obey
\begin{equation}
n_{\rm scr}(\vec{r})
= -\frac{4\pi\epsilon\epsilon_0}{e^2}\int d^2\vec{r}' \frac{\Delta_{\rm
2D} V_{\rm I}(\r')}{|\vec{r}-\vec{r}'|} + \nu n_0.
\label{eq-quasi-laplace}
\end{equation}
Clearly, the actual charge density is expected to deviate from $n_{\rm
  scr}(\vec{r})$ for several reasons. Firstly, the fluctuations of
$n(\vec{r})$ are restricted as discussed above. Secondly,
(\ref{eq-quasi-laplace}) is valid for the Hartree case only. Taking
the Fock contribution into account will introduce short wavelength
fluctuations due to the tendency for crystallization. Fig.\
\ref{fig-density-int-linear} shows results for the charge density of
interacting electrons at $\nu=0.5$. Broken lines indicate the
regions where $n_{\rm scr}(\vec{r})$ exceeds the range for $\delta
n(\vec{r})$ either below or above, i.e.~areas that cannot be
screened at all and thus exhibit insulating behaviour.
Otherwise, we find the charge density to follow $n_{\rm
scr}(\vec{r})$ very closely. In this regime, the density is well
described by (\ref{eq-quasi-laplace}) and the screening is very
effective. Metallic regions dominate over insulating ones and
transport is expected to be good.

We hence have established that interactions at the level of HF
already give rise to substantial differences in the LDOS and local
charge density distributions. Thus one might wonder why theoretical
models based on non-interacting charges work at all. However,
comparing Figs.\ \ref{fig-density-non} and
\ref{fig-density-int-linear} side by side, we see that the charge
density, following the percolating equipotential at half filling for
the non-interacting Fig.\ \ref{fig-density-non}, retains many of the
properties of the HF-interacting charge density line given by
$n_{\rm scr}(\vec{r}) = \bar{n}_{\rm scr} = 0$. Thus non-interacting
percolation-type arguments for network models
\cite{ChaC88,CaiR05,KraOK05} should indeed provide a good starting
point for a qualitative description of the QH transition.

\section{Compressibilities and Coulomb blockade}
\label{sec-coulomb}

In order to understand the experimental results
\cite{IlaMTS04,MarIVS04} we need to focus on the physics of the 2DES
near the Landau band edges. As opposed to the band centre, in this
regime the measured compressibility exhibits very strong
fluctuations as a function of electron density which is due to sharp
jumps in the chemical potential upon varying the carrier density. A
clearer picture of what the microscopic situation is can be obtained
by a spatially resolved scan of $\kappa^{-1}$. Unlike the
fixed-position $(B,n_{\rm e})$-scans, the SET tip will thereby pick
up the spatial variations of the chemical potential change. In the
measurements of Ilani and coworkers \cite{IlaMTS04}, spatially
localised jumps of the chemical potential are clearly identifiable
which form bent segments within the $(x,n_{\rm e})$-plane. These
jumps are fingerprints of local charge accumulation and allow to
conclude that quantum-dot-like structures of electrons (near an
empty Landau band) or holes (near a full Landau band) must be formed
within the sample. The multitude of lines in the compressibility
originating from the dots are indications of Coulomb blockade
effects and provide information on the position, charging as well as
spatial extent of the dots. Since the charge is confined to a small
region within the sample, Coulomb repulsion amongst the particles
requires a comparably large amount of energy for adding or removing
an electron to or from the dot. An increase in the number of
electrons is thus accompanied by equidistant jumps in the chemical
potential. The bending of the compressibility patterns is a feature
of the SET tip bias which acts as an additional potential at the tip
site. As the tip scans across a particular dot, it affects its
charging condition as a function of distance to the centre which
results in the observed arc-like distortion.

\begin{figure}[tbhp]
  \center
  \includegraphics[width=\figwidth]{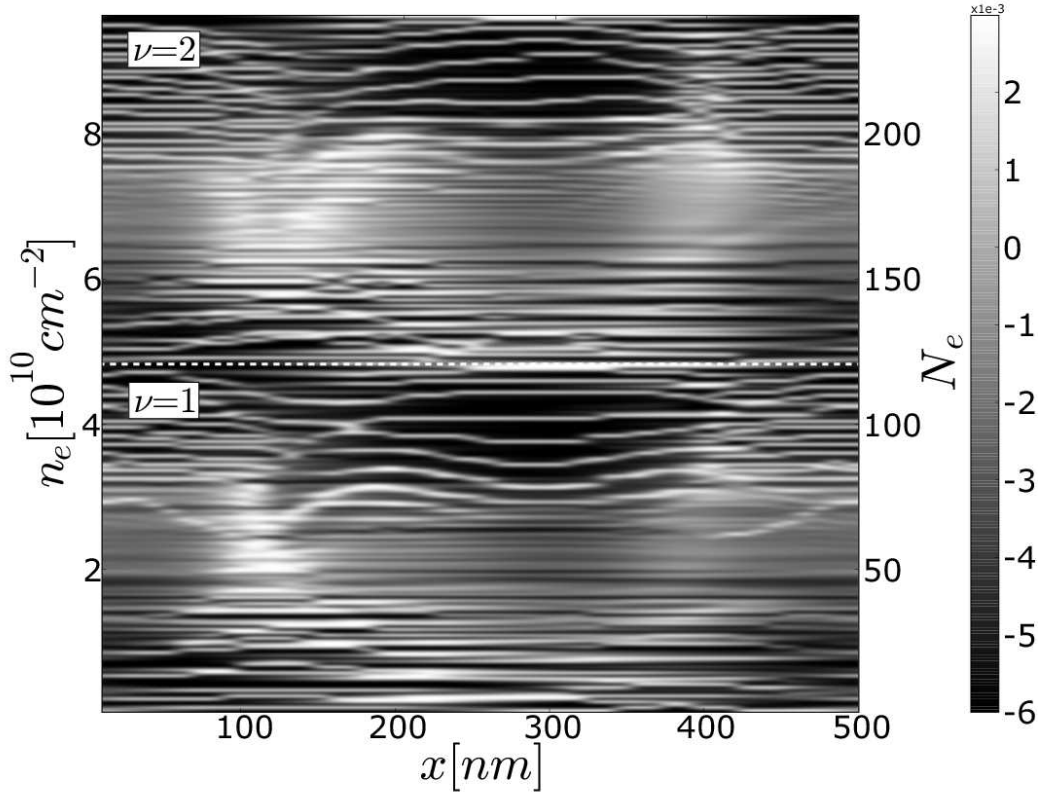}
  \caption{Inverse electronic compressibility $\kappa^{-1}$ for a
    HF-interacting system of size $L=500$nm as a function of position
    $x$ and electron density $n_{\rm e}$ in the two
    lowest Landau levels without spin. The $y$ coordinate has been
    fixed at $250$nm. The patterns correspond to jumps in the chemical
    potential due to charging of local charge puddles (see text for
    details).}
  \label{fig-coulomb}
\end{figure}

Now we want to investigate whether our model is capable of producing
dot spectra as found in the experiment. The change of the local
chemical potential with respect to the electron density can be
computed for our system by noting that the local chemical potential
is the functional derivative of the total energy with respect to the
spatial electron density
\begin{equation}
\mu(\vec{r}) = \frac{\delta E_{\rm tot}[n]}{\delta n(\vec{r})} =
\frac{\delta}{\delta n(\vec{r})}
\iint\dr\dr'\frac{n(\r)n(\r')}{|\r-\r'|} =
\int\dr'\frac{n(\r')}{|\r-\r'|}~.
\end{equation}
Hence, we confirm the anticipated results that the chemical
potential is simply the electrostatic potential due to the electrons
in the 2DES. In terms of our previous definitions, the electrostatic
potential of the 2DES reads
\begin{equation}
\phi(\vec{r}) = L^{-2}\sum_\sigma \sum_{n,k,n',k'}\sum_{\vec{q}}
\VI(\vec{q})\mathbf{D}^{\sigma}_{n,k;n',k'} S_{n,k;n',k'}(\vec{q})
   \exp(-\i\q\r)
\end{equation}
and the local inverse electronic compressibility can be evaluated as
\begin{align}
\kappa^{-1}(\r) \propto \mu^{N+1}(\r) - \mu^{N}(\r) \propto
\phi^{N+1}(\r) - \phi^{N}(\r).
\end{align}
We have calculated $\kappa^{-1}(\vec{r})$ for the lowest two Landau
levels for a system of size $L=500$nm without spin as a function of
position and carrier density at magnetic field $B=2$T. The SET tip
potential, $V_{\rm Tip}(\vec{r})$, has been approximated by a
Gaussian function
$
V_{\rm Tip}(\vec{r}) = {w_{\rm Tip}}/{\pi
  d_{\rm Tip}^2} \exp[{-(\vec{r}-\vec{R}_{\rm Tip})^2/d_{\rm Tip}^2}]
$ 
with $w_{\rm Tip} = 6$meV and $d_{\rm Tip} = 5$nm. The sample is
then "probed" along the $x$-axis at $\vec{R}_{\rm Tip}=(i/50) L$,
where $i = 1,\dots,50$. For each position of the tip the total
potential has to be evaluated and a complete HF run is carried out
since the tip affects the electron density at each position
differently. The results are presented in Fig.\  \ref{fig-coulomb}.
We find remarkable agreement with the experimental results. Distinct
charging patterns appear at the Landau band edges, whereas in the
centre of the bands the features are much weaker. Thus, we have
shown that our model is able to exhibit Coulomb blockade
patterns. We consequently also expect the $(B,n_{\rm
e})$-calculations to exhibit charging patterns similar to the
experimental results and indeed in Ref.\ \cite{SohR07} we show
good agreement between our model and the results of Ref.\ \cite{IlaMTS04,MarIVS04}.

\section{Interaction effects in the Hall conductivity}
\label{sec-interaction}

The compressibility patterns found in the experiments discussed
above were preceded by transport experiments that revealed patters
in the conductance, also clearly interaction mediated. Cobden et
al.~\cite{CobBF99} reported on Hall conductance measurements on
mesoscopic MOSFET devices in the $(V_{\rm g},B)$-plane, where
$V_{\rm g}$ is the gate voltage, that displayed a contradictive
behaviour to the expectations deduced from a single-particle model.
The gate voltage $V_{\rm g}$ can be linked to the electron number
density as $dn_{\rm e}/dV_{\rm g} = C/e $, where $C$ is a material
specific constant. A line in the $(V_{\rm g},B)$-plane originating
from $V_{\rm g}(n_{\rm e}=0)$ can thus be labeled by a certain
filling factor $\nu$.

The fluctuations around the plateau transitions found in Ref.\ \cite{CobBF99} at
half integer filling
factor are apparently correlated over a large interval
of the magnetic field. Surprisingly, however, the lines formed
by the conductance extrema in the $(V_{\rm g},B)$-plane
evidently align with {\it integer} filling factors. This behaviour
conflicts with what is expected from a single particle model
\cite{JaiK88}, where extrema in the transport coefficients
can be linked to resonances in the disorder potential.

Cobden et al.~\cite{CobBF99} suggested a scenario based on dividing
the charge density in the sample into compressible and
incompressible regions. Wherever the density is free to fluctuate,
i.e.~where the local filling $2\pi l_c^2n_{\rm e}(\r)$ lies well
away from an integer value, the 2DES exhibits metallic behaviour.
Wherever the density fills a whole Land level, i.e.~ $2\pi
l_c^2n_{\rm e}(\r)$ is close to an integer, the density is
incompressible and thus insulating. In the spatial density profile,
the metallic regions form puddles that are enclosed by insulating
boundaries where the density corresponds to a full Landau level.
With electron-electron interactions present, transport through the
sample is now strongly influenced by those metallic puddles that are
always surrounded by an incompressible density strip. The
conductance peaks can then be associated with the charging condition
of the puddles and therefore with the shape of the puddles. It is
reasonable to assume that a particular density profile does not
change along lines of constant filling factor in the $(V_{\rm
g},B)$-plane, only the average density as $n_{\rm e} = \nu eB/h$.
Thus it is clear that along lines where $\nu$ is an integer, the
shape of the incompressible puddles (contours where $2\pi
l_c^2n_{\rm e}(\r)$ is an integer) remains roughly constant,
therefore its charging condition and also the conductance extrema.

\begin{figure}[tbhp]
  \center
  \includegraphics[width=\figwidth]{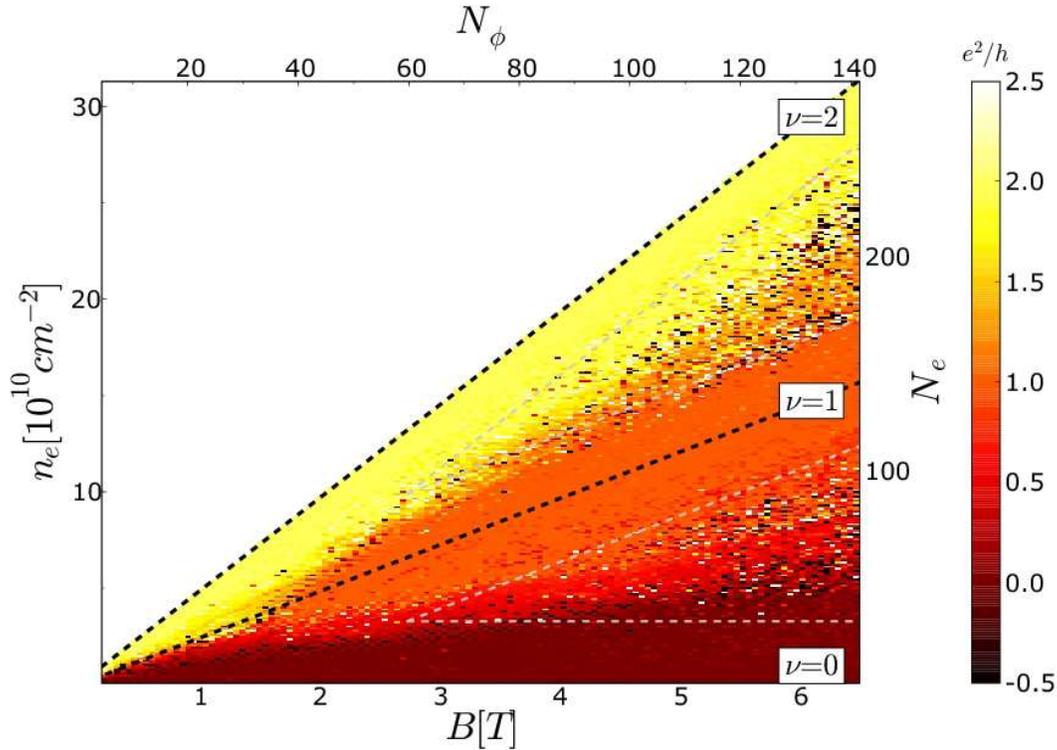}
  \caption
   [$\sigma_{xy}$ for the lowest two spin levels of a HF-interacting
    system with disorder strength $W/d^2 = 2.5$meV in the $(B,n_{\rm
e})$-plane]
   {Hall conductivity, $\sigma_{xy}$, for the lowest two {\it spin
levels} of a HF-interacting system of size $L=300$nm with disorder
strength $W/d^2 = 2.5$meV in the $(B,n_{\rm e})$-plane. The two spin
levels are well separated by virtue of the exchange enhancement of the
spin splitting. Integer filling factors are indicated by the black
broken lines, whereas the grey broken lines indicate boundaries between
linear and non-linear screening \cite{SohR07}.}
  \label{fig-sxy2d-v4-c03-spin}
\end{figure}

Using the Kubo formula for the conductivity \cite{Kub57,SohR07b}, we
have calculated the Hall conductance in the $(B,n_{\rm e})$-plane
for different disorder and interaction strengths.
In Fig.\  \ref{fig-sxy2d-v4-c03-spin} we show the lowest
two spin levels of a system of {\em HF-interacting} electrons of size $L=300$nm. The disorder strength is chosen as $W/d^2=2.5$meV
with an impurity range of $d=40$nm. The spin levels are well separated by the exchange enhanced spin
splitting. We observe stable plateaus around integer filling factors,
which are indicated by the black broken lines.
We have also indicated the boundary between the non-linear and the linear
screening regime by grey broken lines \cite{SohR07}.

From our numerical results, we observe wide, stable plateaus
at the integer filling factors.
In contrast to a single-particle calculation \cite{SohR07b} where
the widths of the plateaus depend linearly on the magnetic field $B$,
the widths of the plateaus remain constant as a function of $B$
when electron-electron interactions are taken into account.
Similarly to the
compressibility calculations, the estimation of the cross-over
between the non-linear and linear screening appears to describe the
competition between disorder and interactions well. The constant
width of the plateau is in agreement with the experimental finding
of Ref.\ \cite{CobBF99}. The expected alignment of conductance peaks
along integer filling factor is, however, absent in our
calculations. Instead, we observe seemingly random conductance jumps
in the centre of the bands. We attribute this behaviour to the
strong exchange correlation in this regime, which was also
predominant in Ref.\ \cite{SohR07}, where the compressibility became
strongly negative in the centre of the bands. Therefore we conclude
that exchange induced effects dominate over charging and Coulomb
blockade effects.

\begin{figure}[tbh]
  \center
  \includegraphics[width=\figwidth]{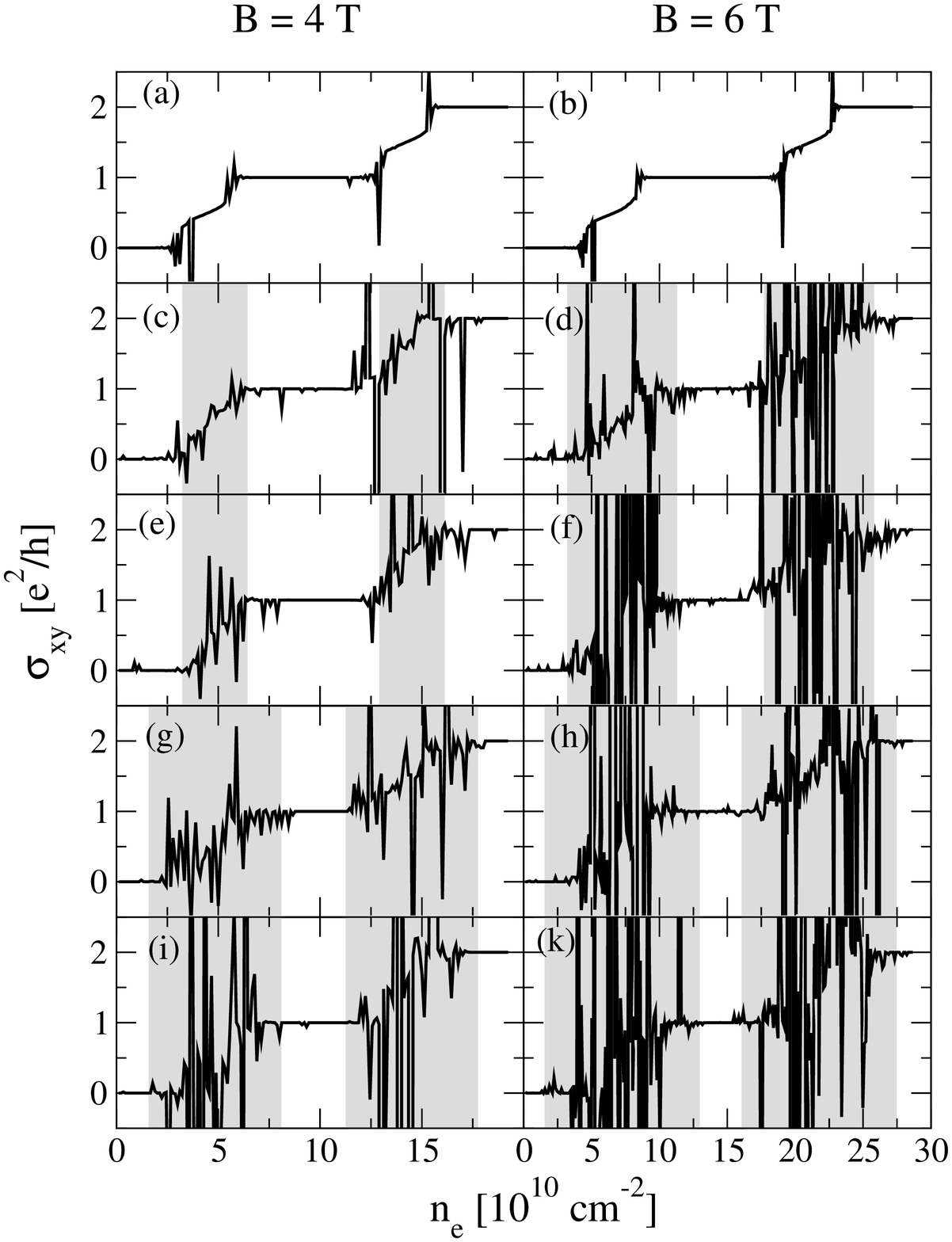}
  \caption
   {Hall conductivity, $\sigma_{xy}$, as cross-sectional cuts for $B=4$ (left column) and $6$ (right column) and (a,b) two spin-less Landau levels with no Coulomb interaction, (c,d) and (g,h) a spin-split Landau level at $W/d^2 = 2.5$meV and $W/d^2 = 1.25$meV, respectively. Panels (e,h) and (i,k) refer to two orbital, spin-less Landau levels at $W/d^2 = 2.5$meV and $W/d^2 = 1.25$meV, respectively. The grey areas indicate the  linear screening regime \cite{SohR07} which widens for weaker disorder (bottom two rows).
   }
  \label{fig-sxy1d}
\end{figure}

In Fig.\  \ref{fig-sxy1d} we show cross-sectional cuts in the $(B,n_{\rm e})$-plane at $B=4$T and $6$T.
Results for 2 Landau levels of non-interacting (a,b), strongly
disordered (c,d,e,f) and weakly disordered (g,h,i,k) systems are
depicted, corresponding to $2$ spin-less Landau levels (c,d,g,h) or
$1$ spin-split Landau level (e,f,i,k).
We have indicated the linear screening by the vertical grey shading.
The plateau regions of localised electrons align well with the
estimation formula of Ref.\ \cite{SohR07} and, to the accuracy of
this simulation, their width is indeed independent of the magnetic
field. The delocalised regime of the plateau-to-plateau transitions
on the other hand increases strongly. This is in contrast to the
single particle result, where both, the plateau and the transition
region increases with the magnetic field. Whereas the
compressibility stripes are fingerprints of the non-linear
screening, i.e.~the insulating regime, the increasing width of the
plateau transition demonstrates an interaction promoted
delocalisation. We regard this behaviour as another important
manifestation of the competing interplay between disorder and
electron-electron interactions even in the integer QH effect.

\section{Conclusions}
\label{sec-conclusions}

We have investigated numerically how the interplay
of electron-electron interactions and disorder
affects transport and thermodynamic properties of electrons in the IQHE. We diagonalised the Hamiltonian for electrons confined to two dimensions and subject to
a perpendicular magnetic field
in the suitable basis of Landau functions and treated
interactions in an effective, self-consistent HF
mean-field approximation. Our results are qualitatively
in very good agreement with recent experiments. More
quantitative statements especially regarding spin-splitting
enhancement might be limited by the unscreened HF
approach used in this paper. Electron correlations beyond the
exchange interaction may lead to a screening of the
exchange term. However, a simple replacement of the exchange
term by a screened one will destroy the self-interaction
cancelation occuring among Hartee and Fock term, thereby
introducing a new source of error. We believe that
for the purpose of this investigation our approach
describes the important physical effects well \cite{YanWM01,MacOL86},
especially for thermodynamic quantities where the system is given enough
time to relax into equilibrium. The possibility to go
beyond a simple electrostatic theory \cite{ChkSG92}
by incorporating the exchange correlation effects, such as
crystalisation and enhanced spin band separation,
while at the same time allowing for the treatment of
system sizes comparable to experiments, render
the HF method a suitable choice.
In that respect more complicated approaches such as density functional
schemes \cite{IhnZ06} may not outweigh the advantages of the
unscreened HF approach.

We close with some final remarks. Whereas disorder is usually associated with
a reduction of signal quality,
in the IQHE the concurrence of a magnetic field,
reduced dimensionality, and disorder leads to
a remarkable resilience of quantised transport.
While the critical behaviour at the transition
appears unaffected, localisation in the band tails
is changed by interactions, yielding a significant
change in the widths of the plateau regions.
Although single-particle
models can well describe many aspects
of quantum Hall physics, only
by taking interactions into account can the whole spectrum of
experimentally observed features
be satisfactorily understood.

\begin{acknowledgement}
  We thankfully acknowledge discussions with J.\ Chalker, N.\ Cooper, A.\ Croy, N.\ d'Ambrumenil, K.\ Hashimoto,
B.\ Huckestein, M.\ Morgenstern, J.\ Oswald, and M.\ Schreiber. This work has been supported partially by the Deutsche Forschungsgemeinschaft via Schwerpunktprogramm ``Quantum-Hall Systeme'' (Ro 1165-1/2/P).
\end{acknowledgement}

\bibliographystyle{prsty}

\end{document}